\documentstyle[aps,multicol,prb,graphicx]{revtex}
\begin{document}

\title{Flexible Linear Polyelectrolytes in Multivalent Salt Solutions: Solubility
Conditions}
\author{F. J. Solis and M. Olvera de la Cruz \\
Northwestern University, Department of Materials Science and
Engineering\\ Evanston, IL 60208, USA \\
e-mail:f-solis@northwestern.edu, m-olvera@northwestern.edu}
\maketitle

\maketitle
\begin{abstract}
Polyelectrolytes such as single and double stranded DNA and many
synthetic polymers undergo two structural transitions upon
increasing the concentration of multivalent salt or molecules.
First, the expanded-stretched chains in low monovalent salt
solutions collapse into nearly neutral compact structures when the
density of multivalent salt approaches that of the monomers. With
further addition of multivalent salt the chains redissolve
acquiring expanded-coiled conformations. We study the
redissolution transition using a two state model [F. Solis and M.
Olvera de la Cruz, {\it J. Chem. Phys.} {\bf 112 } (2000) 2030].
The redissolution occurs when there is a high degree of screening
of the electrostatic interactions between monomers, thus reducing
the energy of the expanded state. The transition is determined by
the chemical potential of the multivalent ions in the solution
$\mu$ and the inverse screening length $\kappa$. The transition
point also depends on the charge distribution along the chain but
is almost independent of the molecular weight and degree of
flexibility of the polyelectrolytes. We generate a diagram of
$\mu$ versus $\kappa^2$ where we find two regions of expanded
conformations, one with charged chains and other with overcharged
(inverted charge) chains, separated by a collapsed nearly neutral
conformation region. The collapse and redissolution transitions
occur when the trajectory of the properties of the salt crosses
the boundaries between these regions. We find that in most cases
the redissolution occurs within the same expanded branch from
which the chain precipitates.
\end{abstract}
\begin{multicols}{2}
\section{Introduction}
The precipitation and dissolution of linear polyelectrolytes with
the addition of multivalent salt or particles have been
extensively studied
\cite{widom,raspaud,raspaud2,olvera,joanny2,bloom,previous,rouzina}.
In DNA the precipitation provides a promising mechanism to
''pack'' long DNA, a critical problem in gene therapy. Moreover,
it is correlated with highly accelerated rates of DNA renaturation
and cyclisation\cite{siko1,siko2}. Therefore, it is important to
determine the concentration of multivalent salt or particles at
which the precipitated DNA dissolves in the solution. Furthermore,
it is crucial to determine the effective charge of the DNA in the
precipitated and dissolved states to determine their interaction
with cells.

It has been shown experimentally that the transitions are nearly
independent of the details of the linear polyelectrolyte (such as
charge density, degree of flexibility, molecular weight, structure function, etc.)%
\cite{raspaud}. That is, flexible (single stranded DNA and
Polystyrene Sulphonate) and semiflexible chains (double stranded
DNA) can be described by the same thermodynamic model even though
flexible chains collapse into amorphous dense spheres and long
semiflexible chains into toroid conformations \cite{gelbart}.

We recently develop a thermodynamic model that explains the
universal nature of the multivalent counterion induced
precipitation transition in low concentration of monovalent
salts\cite{previous}. In this paper we analyze the redissolution
transition by extending our two state model to include large
concentrations of multivalent salt.

\begin{figure}
\includegraphics[totalheight=2.5in,angle=90]{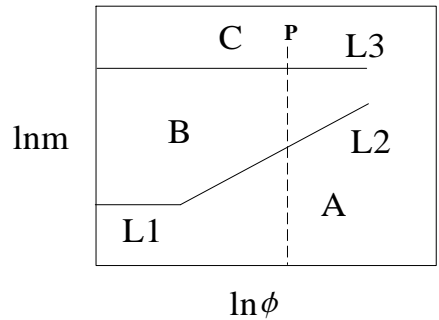}
\caption{Location of the collapse and redissolution transitions in
a diagram of the logarithm of the concentrations of monomers
$\phi$ and of multivalent salt $m$, for dilute, strongly charged
polyelectrolytes. Lines L1 and L2 form the boundary between the
expanded-stretched state and the collapsed state, and line L3 is
the boundary between collapsed and expanded-coiled states. The
line P is a guide used in the text and represents the increase of
multivalent salt at fixed monomer concentration.}
\end{figure}

The experimental diagrams of the logarithm of concentration of monomers $%
\phi $ versus the logarithm of the concentration of multivalent
salt $m$ have three regions (A, B and C) separated by 2
transitions lines (L1-L2 and L3) \cite{raspaud,raspaud2}, shown
schematically in Fig. 1. Region A corresponds to polyelectrolytes
dissolved in water with expanded-stretched conformation at low $m$
values. Region B at intermediate multivalent salt concentrations
$m$ is a solution of collapsed polyelectrolytes (sometimes
considered as a coexistence of two phases: one rich and one poor
in polyelectrolytes). Region C at large $m$ values contains
polymers dissolved in water in expanded-coiled conformations. The
transition line between the A and B regions, L2,
at low monovalent salt concentrations is linear with a slope comparable to $%
1/z$, where $z$ is the valence of the counterions of the added
salt. In the presence of monovalent salt there is a nearly
horizontal transition line, L1, at very low concentrations of
monomers. The transition line between regions B and C is
horizontal (i.e., independent of the concentration of monomers) in
the full regime. In our previous paper \cite{previous} we
discussed the transition between regions A and B and explained the
transition in the regime L2 as the creation of collapsed
conformations of polymers whose charge is compensated by condensed
multivalent salt. Our model also predicts the transition L1 in the
presence of large amounts of monovalent salt. In this paper we
study the transition between the regions B and C in solutions with
very low (negligible) monovalent salt concentrations.

It is well documented that in polyelectrolyte solutions a fraction
of the counterions condensed along the polyelectrolytes to
decrease their electrostatic energy \cite{manning}. This ion
condensation is crucial to understand the precipitation and
redissolution transitions. In this paper we compute the fraction
of condensed ions as the multivalent salt concentration increases
in the two possible polyelectrolyte conformations: collapsed and
expanded. Our two state model is based on the fact that in salt
free and/or low ionic strength solutions, ion condensation leads
to two possible conformation in linear flexible
polyelelctrolytes\cite{gonzalez}: expanded-stretched with a
reduced effective charge and collapsed nearly neutral with the
cohesive energy of an ionic glass\cite{previous}. Indeed,
multivalent counterions lead to large cohesive energies for nearly
neutral collapsed structures in region B explaining the transition
L2. Since the magnitude of the electrostatic energy of a collapsed
chain in region B is much larger than any entropic energy increase
(due to the decrease of degrees of freedom resulting from the
compaction), our model also describes the precipitation of
semiflexible and rigid-rod polyelectrolytes into toroids and
bundles, respectively, explaining the universality of the
precipitation transition discussed in our previous
work\cite{previous}.

At the precipitation transition we predict that the expanded
chains are slightly charged, and with further addition of
multivalent salt, the collapsed chains (in region B) become
practically neutral, in excellent agreement with recent
electrophoresis experiments \cite{raspaud2}. As a result, the
aggregates interactions in the dilute regime are negligible
justifying a monomolecular collapse model to describe also
multi-molecular aggregation.

The effective charge and size of the expanded conformation (in
region C) changes with further addition of multivalent salt. The
transition L3 occurs at high salt concentrations, and in such
conditions the chains are expected to obey random walk or
self-avoiding walks statistics due to the screening of the
electrostatic interactions \cite{barrat}. Screening is responsible
of the redissolution, as suggested in various models \cite
{olvera,joanny2,rouzina}. We show here that the redissolution is
actually very sensitive to the relation between the chemical
potential of the condensing multivalent particles in the solution
$\mu $ and the inverse screening length $\kappa $, while nearly
insensitive to the degree of flexibility and molecular weight of
the polyelectrolytes. Screening, for example, may be reduced due
to multivalent-monovalent ion associating in the solution,
strongly affecting the effective charge of the expanded chains. We
determine the effective charge of the chains as a function of $\mu $ and $%
\kappa $ including the finite size of the ions and the discrete
nature of the charge distribution along the polyelectrolyte, which
is essential to
compute the electrostatic energies of both the collapsed and expanded states%
\cite{previous}. In a $\mu -\kappa $ diagram the collapsed region
lies between two branches of expanded states: one expanded with a
reduced effective charge and the other expanded with an inverted
charge. In most common situations the chains precipitate and
redissolve within the same branch; i.e., with a reduced effective
charge of the same sign of the bare charge. Our model predicts a
redissolution transition independent of polyelectrolyte
concentration and describes the re-dissolution of flexible,
semiflexible and rigid polyelectrolytes. Indeed, the collapsed and
expanded states discussed here are akin to the multi-molecular
precipitated and dissolved states, respectively, observed in many
polyelectrolytes \cite {raspaud2}, explaining the rather universal
form of the diagram, Fig. 1.

In the Section 2 we summarize the assumptions of our work, and
provide a description of each of the phases of the system. In
section 3 we construct the free energies of two types of
representative states and in Section 4 we give the result of the
free energy minimization. In Section 5 we summarize our results,
compare them with experiments and emphasize the elements of the
theory that have not yet been subject to experimental results. We
end with a brief conclusion in section 6.

\section{Theoretical model}
As in our previous work \cite{previous}, we rely in a two state
description of the system. Instead of considering all possible
states of the system (describing them, for example by a changing
scaling exponent), we simply assume that the minimum lies in one
of two extremes: collapsed or expanded conformations. The expanded
conformation, however, is a function of the concentration of salt
in the solution. For example, at low monovalent salt
concentrations the chains are expanded rods, and at large
concentrations they are expanded coils \cite{joanny2}. The free
counterions and solvent can be integrated out provided we take
into account the electrostatic interactions between the
non-condensed counterions in the solution.

The condensed (collapsed) state is treated as an ionic glass
\cite{previous} so that its energy can be approximately calculated
by techniques from solid state physics \cite{wigner}; i.e, we
assume that fluctuations play no important role and can be
neglected. For a very large chain, at low enough temperatures, the
bulk of the collapsed state should acquire an almost crystalline
structure. The finite size of the chain, the connectivity of the
monomers, and finite temperature effects must induce defects into
the structure and the local structure should be glass-like, where
the perfect order has not been achieved. An ionic glass structure
was indeed observed in recent numerical studies of
polyelectrolytes in bad solvents \cite{thiru}.

In both states of the systems, the collapsed and expanded states,
it is necessary to keep track of the finite size of the
counterions and monomers. This is clear in the case of the
collapsed state, but even in the expanded states there is an
important contribution arising from interactions with the nearest
neighbors, and this energy can only be calculated by considering
the finite size of the particles \cite{previous}. For example, a
naive coarse-grained model that takes into account only effective
charges would assign a zero energy to a neutral cluster, while it
can actually have a very strong cohesive energy. The redissolution
into an expanded state occurs at high salt concentrations and the
effective interactions between monomers are screened. In these
conditions the energy associated with interactions between far
away segments of the chain can be treated as a small perturbation
with respect to that of the neutral polymer state.

The properties of the salt at different concentrations are
summarized into two parameters: an effective chemical potential
$\mu $, and an inverse screening length $\kappa $. We will present
our results for general values of these parameters and in
particular consider the case were the parameters are related by a
Debye-Huckel law. We also discuss the effect of
multivalent-monovalent ion association, the Bjerrum model
\cite{book}, in the diagram.

Let us consider the series of states of a single polymer chain
along a line of constant polymer concentration (line P in Fig. 1)
with increasing multivalent salt concentration. We start in a
state in region A with few counterions condensed. The condensed
counterions lie very close to the chain, and cannot be considered
as mixed in the solvent. The effective charge induces repulsions
between the segments of the chain, and the chain acquires extra
stiffness, making it into a rod-like structure.

When a small quantity of multivalent salt is added to the system,
it dissociates, and the multivalent counterions are absorbed to
the chain because the electrostatic energy to be gained from the
condensation is larger (per unit charge) than that of the
monovalent counterions. Furthermore, the entropic loss due to
condensation is lower for the multivalent counterions. When the
charge of the added salt roughly equals the bare charge of the
polymers the effective charge of the chains quickly goes to zero.
At this point most charges of the chain are compensated by the
condensed multivalent counterions and a few monovalent ones. Since
a group of charges with total charge zero has a minimal energy
when they are set in a compact configuration, a collapsed
conformation is acquired. This transition is of course also
possible for the case of pure monovalent salt, but it requires
either larger concentrations or lower temperatures to offset the
entropic contributions. In other words, the determining factor for
the transition is the strength of the electrostatic interaction,
and the introduction of multivalent salt acts as a jump in
parameter space and is not equivalent to a smooth increase in
monovalent salt \cite{previous}.

In the collapsed state the absence of uncompensated charges in the
polymer reduces its solubility in water and polar solvents. It is
not clear from experiments, however, whether the collapsed
flexible polymers aggregate into large structures or if they form
a dense solution of individually collapsed chains. Though we
assume monomolecular collapse, the predicted transition points
should not exhibit strong dependence on the number of chains
aggregated since if the collapsed state is neutral, the residual
interactions between chains (promoting or opposing multimolecular
aggregates) are energetically small.

Since the collapsed structure is almost neutral, the increase in
salt concentrations has little effect in the absolute value of the
energy of the chain. The change in the environment affects only
the interaction between the positive or negative small excess (or
defect) of charge in the polymer, which would be located at the
surface of the collapsed structure. The energy of the charges
packed inside the structure is dominated by the interaction with
their nearest neighbors and is not subject to screening (there are
no floating ions between them). On the other hand, the energy of
an expanded chain with non-zero effective charge will be greatly
reduced by the increased screening.

For the purposes of comparison with the collapsed structures, we
can consider a very rigid chain that cannot be collapsed into a
compact structure. We can ignore for the moment the fact that this
chain would likely form a bundle with other chains, and we can
assume that we have isolated it by perhaps mechanical means. At
higher multivalent salt concentrations, the rigid polymer
continues absorbing ions, and thus it can become neutral and even
overcharged (effective charge with opposite sign to the bare
charge, as it occur in many systems \cite
{bruinsma,pincus2,safinya,levin,eugene}). With further increase in
the amount of salt the trend can be reversed, and the chain can
again become, first neutral, and then simply charged (same charge
sign as the bare charge). The increase of salt concentration
reduces the entropic penalty for the condensation of counterions.
The counterions continue condensing into the polymer as long as
there is a reduction in electrostatic energy due to the
condensation. The increase in condensation stops when the typical
screening length of the free counterions is of the order of the
monomer separation, the free counterions can reduce their energy
by associating with each other, and a fraction of them will do so
instead of associating with the polymer.

At large salt concentrations, the chain reverts from collapsed to
expanded, giving the transition $B-C$. In the expanded state the
chain needs not to be neutral. The screening of electrostatic
interaction under these conditions strongly reduces the
interaction between the uncompensated charge of the polymer, and
it is clear that the repulsions will not be strong enough to
create a rod-like state. In these conditions an expanded coil
state is more likely. In terms of the effective charge of the
polymer we find that two scenarios are in principle possible. If
the chemical potential of the counterions in the solvent is low
enough while the screening length is reduced, the expanded state
is overcharged since there is an effective decrease of energy by
increasing the condensation to the chain, as recently discussed by
Nguyen et al. \cite{rouzina}. On the other hand, when the
screening length is small, it is possible that the interactions
between coions and counterions in the solvent become important,
and it is much harder for the chain to acquire the counterions.
This is, the chemical potential for the extraction of counterions
from the solvent is large and therefore the effective charge at
the redissolution point has the same sign as the bare charge. We
show in the next section that though there is a possibility for
overcharged polyelectrolytes at the redissolution, in the sense
that this state can be preferred to a collapsed one, the salt
properties required for this situation ($\mu $, $\kappa $) are
unlikely to be realized in practice for water soluble multivalent
ions.

\section{Calculation of free energies}
We assumed that the polymers are monodisperse with monomer number
$N$. Each monomer is charged positively, and is monovalent. The
original counterions of the polymer also have valence $1$. The
polymer density is $\phi $. We measure all distances with respect
to the monomer size $b$ (so that $b=1$ in our units). It is
necessary for more precise calculations to also give more detailed
geometrical information such as the aspect ratio of the monomers
and the respective size of all the ions. Here we only need to
further consider, besides the size of the monomer, the radius of
the multivalent counterions $a$. The monovalent salt concentration
$s$ is set equal to zero, but there are monovalent counterions
originally dissociated from the polymer with density $\phi $, and
the multivalent salt, with valences $z:1$ has concentration $m$.
All energies are measured in units of $k_{B}T$, (the Boltzmann
constant times the temperature), and consider only energies per
monomer. We form the dimensionless Bjerrum number $B$, equivalent
to the Manning parameter, as
\begin{equation}
B=\frac{e^{2}}{4\varepsilon _{0}\pi bk_{B}T},
\end{equation}
where $\varepsilon _{0}$ is the permitivity of the solvent
(water). The typical value of this number at room temperature in
strongly charged polyelectrolytes is larger than $1$ and as high
as $4.2$ for double stranded DNA. We assume here for simplicity
that $B$ is a constant, independent of the number of locally
associated monomers and ions. Given that in both states water is
always present, the corrections in the values of $\epsilon $ are
expected to be minor \cite{losrusos}.

In both expanded and collapsed states we describe the system by a
particular geometry and a given fraction $x$ of condensed charge.
The free energy is separated into a contribution from the solvent
and free ions, and another part from the interaction of the
monomers and charges condensed to them. The fraction condensed is
obtained self-consistently from minimization over all possible
condensed fractions. The form of the entropic term is the same in
both cases, but it is easier to consider the electrostatic
contributions separately.

\subsection{Free counterions contributions}
It is convenient to separate the region occupied by the polymer
from the rest of the solvent. To calculate the free energy of the
solvent region, we first calculate the entropic contribution. We
assume that the charge
condensed to the chain $-xN$ is composed by $x_{s}N$ monovalent ions and $%
x_{m}N/z$ multivalent ions. Thus there are $(s/\phi )+1-x_{s}$
free
monovalent counterions per monomer, and thus the energy per monomer in $%
k_{B}T$ units is
\begin{equation}
F_{s1}=(\frac{m}{\phi }-\frac{x_{m}}{z})\ln (m-x_{m}\frac{\phi }{z})+(\frac{s%
}{\phi }+1-x_{s})\ln (s+(1-x_{s})\phi ).
\end{equation}
In this expression we omitted the contribution of the ions with
the same charge of the monomers, since it is assumed that they do
not condense, and thus have a constant contribution. We have also
omitted the contribution of the solvent, since the concentration
of the salts is small compared to the solvent. It is possible to
add in this expression an interaction term between the solvent and
the ions, but for simplicity we neglect these corrections here;
its effect would be only to shift the position of the L3 line in
the phase diagram. It is important to consider, however, the
effective free energy of interaction between all free ions. A
simple way to do this is to use the modified Debye-Huckel form:
\begin{equation}
F_{s2}=\frac{1}{4\pi \phi }\left( ln(1+\kappa a)-\kappa
a+\frac{1}{2}(\kappa a)^{2}\right) .
\end{equation}
In this expression $a$ is the hard-core radius of the salt ions
and $\kappa $ is the inverse screening length produced by the
salt:
\begin{equation}
\kappa ^{2}=4\pi B(z^{2}(m-x_{m}\phi /z)+zm+(2s+(1-x_{s})\phi )).
\end{equation}
In the very dilute polymer limit, we can simplify these
expressions, and use instead
\begin{equation}
\kappa ^{2}=4\pi z(z+1)Bm,
\end{equation}
and expand the total free energy of the solvent
$F_{s}=F_{s1}+F_{s2}$ in a series in the overcharge fraction
$y=x-1$. The constant term is irrelevant and we retain only the
linear term:
\begin{equation}
F_{s}=(\partial _{y}F_{s})y=\mu y,
\end{equation}
where $\mu $ is now the effective chemical potential for the
extraction of ions from the solvent into the condensed region.
Using the same high salt limit as before we obtain:
\[
\mu =-\frac{1}{z}\ln m+\frac{1}{8\pi (z+1)m}\frac{\kappa
^{3}}{1+\kappa a}.
\]

\subsection{Expanded state energy}
In our previous work \cite{previous}, we considered only the
energy of rod-like conformations. Since our interest is the regime
with large amounts of salt, the screening strongly reduces the
rigidity of the chain due to excess charge. It is then adequate to
consider a Gaussian conformation (self-avoiding walk conformations
give very similar results, as it can be seen from the functional
form of the free energy computed below). The electrostatic
contribution to the free energy is calculated as a function of the
condensed charge, and we set the neutral state $x=1$ as reference.
The free energy is expressed as the sum of three terms. A constant
term reflects the energy of the neutral state, a term linear in
the excess charge $y=(x-1)$ appears due to addition or extraction
of a charge into or from the neutral state. Finally, the
interaction between segments of the chain with excess charge gives
rise to a term quadratic in $y$. We write this as:
\begin{equation}
F_{e}^{e}=g_{0}^{e}+g_{1}^{e}y+g_{2}^{e}y^{2}.
\end{equation}
A coarse-grained model clearly gives rise to the last term, and
the first two can be understood, in that context, as
regularizations that take into account the finite size of the
particles involved. In the limit of large amounts of salt-added,
most of the condensed charge comes from multivalent ions, and
thus, we will take $x=x_{m}$, and $x_{s}=0$. The terms are
graphically represented in Fig.2.

 For the calculation of the energy of the neutral state we
basically apply the Wigner principle\cite{wigner}: in a dense
ionic system, the free energy can be approximated by the
contribution of nearest neighbors that effectively cancel the
charge of the particle. In the case of the collapsed state, below,
it is more suitable to consider an approximation based on the
calculation related to a full infinite lattice.

\begin{figure}[tbp]
\includegraphics*[totalheight=2.5in,angle=90]{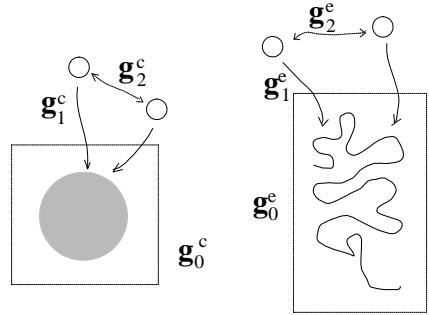}
\caption{Graphical representation of the terms required to
calculate the energy of the system in both states. The $g_{0}$
term measures the energy per monomer of the chain in a neutral
state. $g_{1}$ measures the energy of adding an additional charge
to the neutral conformation. $g_{2}$ is the energy of interaction
between added charges and is obtained by a coarse graining
procedure. All the excess charge in the collapsed state (b) is
assumed to lie on its surface.}
\end{figure}

 The basic neutral
cluster of the linear polyelectrolyte consists of one multivalent
counterion, and $z$ monomers. Since all particles in the cluster
(with charges $q_{I}$, both monomers at distances $r_{ij}$ from
each other, all roughly equal to the monomer size $r=b=1$, the
electrostatic energy of the cluster is
\begin{equation}
E=\sum_{i\neq j}B\frac{q_{i}q_{j}}{r_{ij}}=-B(z^{2}+z(z-1)/2)
\end{equation}
Dividing by $z$ monomers, we obtain:
\begin{equation}
g_{0}^{e}=-B(z+1)/2.
\end{equation}
When an extra multivalent ion is added to this cluster, the
counterion can have roughly the same interaction with the chain
monomers, but can locate itself away of the first counterion, say,
by putting a monomer in between them so that $r\approx 2$. Thus we
obtain a simple estimate for the energy of the new charged cluster
as:
\begin{equation}
E=B(-2z^{2}+z(z-1)/2+z^{2}/2),
\end{equation}
and
\begin{equation}
g_{1}^{e}=-Bz/2.
\end{equation}

Finally, the two-body term can be calculated assuming a uniform
distribution of charge along the Gaussian chain. A Gaussian chain
has a monomer distribution given, approximately, by $\rho
(r)=1/(2\pi r)$, inside the
volume $V$ limited by the radius of gyration $R$; this is, for $r<R=N^{1/2}$%
. A charge at the center of this distributions interacts with
other charges via the screened potential $Be^{-\kappa r}/r$. The
energy for one such charge is given by the average over the volume
$V$ of this potential:
\begin{equation}
E=\langle V\rangle =\int_{V}d{\bf r}\rho (r)B\frac{e^{-\kappa r}}{r}=\frac{B%
}{\kappa }(1-e^{-\kappa R}),
\end{equation}
and by our previous definition,
\begin{equation}
g_{2}^{e}=\frac{1}{2}\langle V\rangle .
\end{equation}
Notice that while this calculation depends on the assumed Gaussian
conformation, a simple estimate can also be done by using a
general expansion coefficient for the radius of gyration $R\approx
N^{\nu }$, ($\nu =1$ for rod-like, $\nu =2/3$ for self-avoiding
walks, and $\nu =1/2$ for random walks). In such case, we obtain a
scaling for the quadratic precoefficient of the form
$g_{2}^{e}\approx \kappa ^{(1-1/\nu )}$. While these different
models give rise to different estimates, the transition we are
interested in occurs when the screening is of the order of the
size of the monomers, and so, in our units, $\kappa \approx 1$,
and the estimates at that point are not so different.

\subsection{Collapsed state energy}
We repeat a similar approach for the calculation of the free
energy of the collapsed state which is expressed as:
\begin{equation}
F_{e}^{c}=g_{0}^{c}+g_{1}^{c}y+g_{2}^{c}y^{2}.
\end{equation}
Inside the collapsed polymer, we can imagine the environment of a
charge resembling that of a ionic crystal. If the structure of the
crystal is known, the energy per atom in that crystal can be given
in terms of the Madelung constant \cite{feynman},
\begin{equation}
E=-Me^{2}/r=-MB.
\end{equation}
The Madelung constant for an ionic monovalent salt is $M=-1.747$.
There are two important differences in our case. We would like to
consider multivalent ions, which reduce the contribution from
same-charge interactions, and on the other hand the connectivity
of the monomers forces them to be next-neighbors (not the case in
ionic crystals), and thus increases the same-charge interactions.
A way to obtain a rough estimate for $M$ is to consider a simple
geometry in which both monomers and counterions lie in columns
which in turn form a square lattice when a cross section is taken.
The result does not change dramatically when other possible
geometries are considered. As a rule of thumb, we can use an
effective Madelung constant that is larger than unity, but not
much bigger. In the geometry proposed, the number of next
neighbors of counterion is $4z$. Summing the interaction with $4z$
monomers and 4 equally charged next-nearest counterions ($r=2b$)
we obtain the following estimate of the energy per cluster of $1$
counterion and $z$ monomers:
\begin{equation}
E=B(-4z^{2}+z(z-1)+\frac{1}{2}z(4z-z)+\frac{1}{2}\frac{4z^{2}}{2})
\end{equation}
where the interactions with charges outside the cluster are
weighted by $1/2$ to avoid double counting. Dividing again by $z$,
we obtain the coefficient of the energy per monomer:
\begin{equation}
g_{0}^{c}=-B(1+\frac{z}{2}).
\end{equation}
In this approximation we have underestimated the Madelung constant
for $z=1$, for which we obtain $-1.5$ instead of the exact result
$-1.747$, but it is a reasonable approximation and shows that the
energy of this neutral collapsed state is smaller than the neutral
expanded state.

In both the collapsed and expanded states, the addition and
extraction of charges are not completely symmetric processes. It
is energetically unfavorable to add or extract charges from the
collapse bulk state, as shown below in the calculation of
$g_{2}^{c}$. Both processes are however better thought of as
occurring at the surface of the collapsed region. We have also
argued previously \cite{previous}, that while the mobility of
charges in this type of conformations is not so large, the analogy
with a conductor is still useful: all charge excess resides in the
surface. While the local conformation at the surface of the
collapsed state is not equal to the expanded state, it is simpler
to borrow the result for the coefficient related to the addition
of one charge from that case. In both situations the new charge is
put in close contact with a neutral cluster, but sits slightly
closer to the opposite charges. Thus,
\begin{equation}
g_{1}^{c}=-Bz/2.
\end{equation}

Since we assume that the extra charge is sitting at the surface,
the interaction energy can be calculated assuming a uniform
distribution in a shell surrounding the collapsed polymer. We
assume this shell to be spherical with radius $R$, and obtain the
energy per added charge as
\begin{equation}
g_{2}^{c}=E=\frac{1}{2}B\frac{1}{R}=\frac{1}{2}\frac{B}{(3N(1+1/z)/4\pi
)^{1/3}}
\end{equation}
Here we have considered that the $N+N/z$ charged particles are
closely packed and that they occupy each a volume of $b^{3}$.

\section{Comparison between states and Phase diagram}
After the minimization of the free energies of each state with
respect to their respective condensed charge variables, we can
compare each other to find the state of the chain for given
concentration conditions. This minimization can be done directly
from the expressions already given but it is easier to present
simpler formulas if we use a set of approximations described
below. Numerical results presented later reflect these
approximations, but we have checked that numerical minimization of
the full expressions do not lead to important qualitative and
quantitative changes in the results.

The free energy of each of the states, for given concentrations of
monomers and salt correspond to the minima of:
\begin{eqnarray}
F^{e} &=&F_{s}+F_{e}^{e}, \\ F^{c} &=&F_{s}+F_{e}^{c}
\end{eqnarray}
for the expanded and collapsed states respectively. We minimize
now with respect to the overcharge fraction $y$.Taking the first
derivatives with respect to this quantity, we obtain the
(independent) conditions
\begin{eqnarray}
\partial _{y}F^{e} &=&\mu +g_{1}^{e}+2g_{2}^{e}y^{e}=0, \\
\partial _{y}F^{c} &=&\mu +g_{1}^{c}+2g_{2}^{c}y^{c}=0.
\end{eqnarray}

The solutions for the fractions condensed are:
\begin{eqnarray}
y^{e} &=&-(\mu +g_{1}^{e})/2g_{2}^{e}, \\ y^{c} &=&-(\mu
+g_{1}^{c})/2g_{2}^{c}.
\end{eqnarray}
Replacement of this values in the general expressions for the free
energy gives the minimized free energies for each state:
\begin{eqnarray}
F^{e*} &=&F^{e}(y^{e})=\mu +g_{0}^{e}-\frac{(g_{1}^{e}+\mu )^{2}}{4g_{2}^{e}}%
, \\
F^{c*} &=&F^{c}(y^{c})=\mu +g_{0}^{c}-\frac{(g_{1}^{c}+\mu )^{2}}{4g_{2}^{c}}%
.
\end{eqnarray}

For the moment it is convenient to maintain the expressions of the
free energies as functions of the chemical potential instead of
the salt concentration. Of the parameters used to describe the
different states, only $g_{2}^{e}$ has an explicit dependence on
the concentration (through the inverse screening length $\kappa
$)l. Also, it is important to note that the coefficient
$g_{2}^{c}$ is extremely large, and for the values of the chemical
potential that we need to explore, it turns out that the
denominator in the expression for the overcharge is much larger
than the numerator, and the overcharging fraction is very close to
$0$:
\begin{equation}
y^{c}\approx 0.
\end{equation}
This result is clear since we have found that there is a very
strong penalty for any overcharging (or undercharging) in the
collapsed state. The effective charge in a general structure is
affected by the environment (namely, by $\mu $) but the collapsed
structure is only very slightly susceptible to it. Furthermore,
the interaction between the charges in the bulk of the collapsed
structure is not subject to screening by the free particles
outside and its contribution remains constant with changing
chemical potential. Thus, the system splits neatly into collapsed
neutral structures and a homogeneous mixture of solvent and
non-condensed counterions. It is then not surprising that this
state adequately represents what in experiment is close to be two
separate phases.

Consider now the minimum for the expanded state. The coefficient
for charge-charge interactions $g^{e}_{2}$ varies very strongly
with the salt concentration. It remains large when the screening
length is large, since in those conditions all uncompensated
charges interact strongly. On the other hand at large
concentrations of salt the screening drastically reduces the
interaction between segments of the chain.

To obtain an expression for the transition point we subtract the
free energies, and define the energy difference
\begin{equation}
\Delta F=F^{e*}-F^{c*}.
\end{equation}
The system is in the region $B$, when $\Delta F>0$, in region $C$ when $%
\Delta F<0$, and the transition curve is defined by $\Delta F=0$.
Using the fact, discussed above, that the electrostatic energy of
the collapsed state remains essentially constant, we can write the
equation $\Delta F=0$ as
\begin{equation}
g_{0}^{e}-\frac{(g_{1}^{e}+\mu )^{2}}{4g_{2}^{e}}=+g_{0}^{c}.
\end{equation}
One useful way to use and interpret this equation is to consider
it as defining a boundary in the $\mu -\kappa ^{2}$ plane that
corresponds to the properties of the counterions. Within this
diagram, a particular form of the relation between concentration
and screening for a counterion defines a path in the diagram
parametrized by the salt concentration $m$. Examples of this are
presented in Fig.3. for certain polymer properties values, and
different types of $\mu -\kappa ^{2}$ relations.

At this point is very easy to explain the observed near
independence of the
transition on the molecular weight of the chains. Only the coefficients $%
g_{2}$ carry information on the molecular weight, but as we have seen $%
g_{2}^{c}$ drops out of the final equations, and in $g_{2}^{e}$
the molecular weight provides only an exponentially small
correction to the dominant term given by a function of the
screening length.

As shown in Fig.3, and as it can be seen from the fact that the
equations defining the transition are quadratic in $\mu $, the
expanded state splits into two regions, each with non-zero
effective charge, but with opposite signs on each. This diagram is
obtained using the approximations presented above, and a more
precise determination of the lines of transition will involve
consideration of the monovalent counterions, the finite amount of
monomers, and the rod-like state obtained at very low screening.
When these considerations are taken into account, the trajectory
described by the added salt in the $\mu -\kappa ^{2}$ plane will
start within the expanded region, cross into the collapsed region
(giving rise to the L1-L2 line), and continue there until crossing
again, redissolving, into one of the branches with expanded
states. The starting point of the trajectory, with zero
multivalent salt added $m=0$, is located in the branch with
natural charge, at a finite $\kappa $ value (due to the monovalent
counterions), and at $\mu =\infty $. Figure 3 illustrates only the
redissolution.

In Fig.3 we show three different curves defining the properties of
the counterions: a pure Debye-Huckel case where the size of the
counterions are neglected, a modified Debye-Huckel curve with
$a=0.2$ (that is $a=0.2b$), and a third curve that neglects the
interactions between the counterions when they are free considers
only the entropic term. The reversal of direction in the $\mu$
axis of the parametric curve for the Debye-Huckel cases occurs
because of the onset of important favorable interactions between
the free charges. Thus it is clear that the trajectory can end in
the naturally charged branch of the expanded region since there
are more free counterions there, and at the same time the
repulsive intra-chain interactions are small. The overcharged
branch is only accessible if the chemical potential is further
reduced at the same time that the screening increases. For most
cases with sensible sizes for the counterions the reversal in
direction of the parametrized curve comes always before the
redissolution, and thus the redissolution is into the naturally
charged state.

\begin{figure}
\includegraphics*[scale=0.7,viewport= 10 0 300 290]{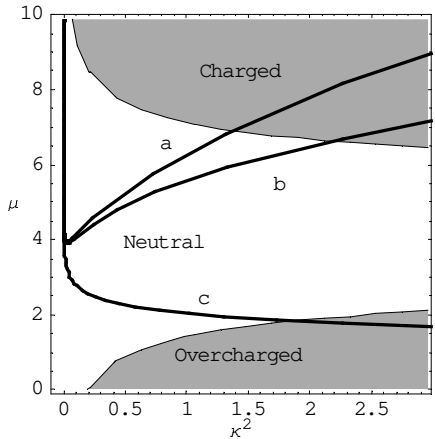}
\caption{Diagram of $\mu-\kappa^2$ showing the regions of expanded
(gray) and collapsed (white) conformations. The solid curves give
the relation of $\mu(\kappa)$ using the Full Debye-Huckel model
for different size of the ions $a$. The first line (a) is for
$a=0$, the second line (b) is for $a=0.2$ and the third curve (c)
is for pure entropy (i.e., neglecting the electrostatic
interactions of the multivalent ions in the solution, and
mathematically equivalent to $a=\infty$). The three lines are
indistinguishable at low concentrations (small values of
$\kappa^2$).}
\end{figure}

It is interesting to note that at high enough salt concentrations
and high valences, the salt might re-associate on its own. In this
case the trajectory in the $\mu -\kappa ^{2}$ plane has a
different shape than that provided by the Debye-Huckel equation.
Indeed, while the results from the Debye-Huckel model are know to
provide good approximations for the free energy of the salt in
solution they do not properly describe the individual dissociated
species. The upturn of the chemical potential essentially
indicates a phase transition for the salt/solvent system. In
reality there is no phase transition but there is a finite a
fraction of ion pairs formed in the solvent (even if they do not
reassociate chemically). We have not presented here the curve in
the $\mu -\kappa ^{2}$ diagram when the association of multivalent
ions in the solution is present, as is modeled, for example, in
the classical Bjerrum model \cite{book}, or more recently by
Fisher, et al.\cite{fisher} This association gives large
corrections to the Debye-Huckel law (for the free energy of the
individual species of ions) and in our case provides a $\mu
-\kappa $ curve that monotonically approaches a horizontal line (a
limiting $\mu $ value), with increasing inverse screening length.
These effects are particularly important for large valence salts
as in our case and we will discuss them in more detail in a future
publication \cite{inprepa}.

\section{Comparison with experimental information}
With the previously obtained solutions we now contrast our results
with the currently known experimental situation. We consider the
shape of the phase diagram, the precise location of the collapse
transition and the predicted structure of the polymers in both of
the phases.

It should be clear that we have already recovered the shape of the
$B$-$C$ transition. Indeed, the (approximate) equation that define
the transition does not have any dependence on the monomer
concentration, and therefore produces a flat line in the $\phi-m$,
or $\phi-\mu$ diagrams. The transition is defined by the choice of
single chains to be in one of two states that must be in
equilibrium with the surrounding environment. This environment,
represented here by means of an effective chemical potential and
screening length, depends strongly on the amount of multivalent
salt and only very slightly on the polymer amount. The more
precise expressions for the chemical potential recovers a small
dependence in the polymer concentration.

There are two important ways in which important changes on the
qualitative form of the diagram can arise.The addition of large
amounts of monovalent salt introduces new features in the phase
diagram, not show in the scheme of Fig. 1. In the introduction we
only mentioned the effects on the $A-B$ transition, but further
addition of monovalent salt, creates a different environment of
the chains, and changes the transition points and the shape of the
transition curve. Secondly, it is clear that as we continue
increasing the amount of polymer in the solution, the interactions
between the chains start to become important and the transition
(if any) is a much more complicated phenomena. From the point of
view of our theory it is clear that the increase in concentration
of monovalent salt and monomers bring about a breakdown of the
assumption that the chemical potential for the overcharging is
dependent only on the amount of multivalent salt. The effective
environment of the chain becomes more complicated and the chemical
potential will now contain important terms coming form the
concentrations of all species and from many-body interactions.

Let us consider two concrete numerical examples. Experiments
performed on polystyrene sulfonate \cite{olvera} at room
temperature with $N=4*10^{3}$,
Bjerrum number of $B=2.87$ ($b=0.5$ nm) and an effective ion radius $a/b=0.2$%
, show a transition line at a concentration of $m=0.2M$ of
$LaCl_{3}$. With these parameters as input, our equation predicts
redissolution at a salt concentration of $m=0.1M$. Secondly,
experiments on double stranded DNA \cite
{raspaud} with very large number of base pairs (we take $N=10^{4})$ and $%
B=4.2$ ($b=.17$ nm)in spermine ($z=4$), $a=1$. obtained redissolution at $%
m=.1M$, while our equations predict a transition at $m=0.04M$.
This compares favorably with the experiments, especially if we
take into account the rough approximations done in the evaluation
of the energies of the system.

According to the theory presented, in phase B, the polymer chains
are almost neutral with a collapsed conformation. In phase C, the
chains are expanded and charged. This suggests, besides other
techniques, to confirm these predictions by means of scattering
experiments that test the structure of the chains, and of osmotic
pressure measurements that can determine the amount of free
counterions in each state (as it has been done for semi-dilute
solutions\cite{delsanti,amis,claudine}). 

\section{Conclusions}
The redissolution transition observed in multivalent induced
precipitated polyelectrolytes with further addition of multivalent
particles or salt was predicted extending a previously developed
two state model\cite {previous} to deal with large salt
concentrations. The electrostatic energies of both, collapsed and
expanded-coiled, states were computed considering the finite size
of the ions and monomers condensed along the chains, and using a
mean field approach for the non-condensed ions. We neglected the
non-ionic short range interactions of the ions with the solvent,
and assume zero monovalent salt.

We found that the redissolution is determined by the properties of
the ionic solution. We calculated the effective charge of the
chains for the different thermodynamic states of the chain.

\subsection*{Acknowledgments}
We thank E. Raspaud and F. Livolant for useful discussions and for
performing electrophoresis experiments \cite{raspaud2}, M. Campos
for computing electrostatic energies of dense finite size systems
of charges, and P. A. Bernikowicz for computing the free energy of
aggregates containing $p=1,2,...$ chains. This work was sponsored
by the National Science Foundation, grants DMR9807601 and
DMR9632472.

\end{multicols}

\begin{thebibliography}{5}
%
\bibitem{widom}  J. Widom, R. L. Baldwin:
             J. Mol. Biol. {\bf 144} (1980) 431 \\
            Biopolymers {\bf 22} (1983) 1595
%
\bibitem{raspaud}  E. Raspaud, M. Olvera de la Cruz,
            J.-L. Sikorav, F. Livolant:
            Biophys. Jour. {\bf 74}, 381 (1998)
%
\bibitem{raspaud2}  E. Raspaud, I. Chaperon, A. Leforestier, F.
        Livolant:
         Biophys. J. {\bf 77} (1999) 1547
%
\bibitem{olvera}  M. Olvera de la Cruz, L. Belloni, M. Delsanti, J. P.
        Dalbiez, O. Spalla, M. Drifford:
        J. Chem. Phys. {\bf 103}, (1995) 5871
%
\bibitem{joanny2}  J. Wittner, A. Johner J. F. Joanny:
             J. Phy. (paris) II {\bf 5} (1995) 635
%
\bibitem{bloom}  V. A. Bloomfield:
     Curr. Opin. Struct. Biol. {\bf 6} (1996) 334
%
\bibitem{previous}  F. J. Solis, M. Olvera de la Cruz:
         J. Chem. Phys. {\bf 112} (2000) 2030
\bibitem{rouzina}  T. T. Nguyen, I. Rouzina, B. I. Shklovskii:
  J. Chem. Phys. {\bf 112} (2000) 2562
%
\bibitem{Livolant}  F. Livolant: Euro. J. Cell Biol. {\bf 33} (1984) 300.
%
\bibitem{siko1}  D. Jary, J. Lal, J.-L. Sikorav:
         Molecular Biology {\bf 321} (1998) 1
%
\bibitem{siko2}  J. L. Sikorav, G. M. Church:
        J. Mol. Bio. {\bf 222} (1991) 1085 \\
        I. Chaperon, J.-L. Sikorav:
         Biopolymers {\bf 46} (1998) 195
%
\bibitem{gelbart}  S. Y. Park, D. Harries, W. M. Gelbart:
         Biophysical J. {\bf 75} (1998) 714
%
\bibitem{gonzalez}  P. Gonzales-Mozuelos, M. Olvera de la Cruz:
     J. Chem. Phys. {\bf 103} (1995) 3145
%
\bibitem{manning}  G.S. Manning:
        J. Phys. Chem. {\bf 88} (1984) 6654
%
%
\bibitem{barrat}  J. L. Barrat, J. F. Joanny:
    Adv. Chem. Phys. {\bf 94} (1996) 16 (1996).
%
\bibitem{thiru}  N. Lee and D. Thirumalai: pre-print cond-mat/9907199
%
\bibitem{wigner}  E. P. Wigner: Trans. Faraday Soc. {\bf 34} (1938) 678
%
\bibitem{book}  N. Bjerrum: Koninklinge Dans. Vidensk. Selsk. {\bf
7} (1926) 9. See: B. E. Conway:  "Ionic Interactions and Activity
Behavior of Electrolyte Solutions" section 7.2.3. in B. E. Conway,
J. O'M. Bockris, E. Yeager: {\it Comprehensive Treatise of
Electrochemistry; v.5.} New York: Plenum Press (1983).
%
\bibitem{pincus2}  E. M. Mateescu, C. Jeppesen, P. Pincus:
         Europhys. Lett. {\bf 46} (1999) 493
%
\bibitem{bruinsma}  S. Y. Park, R. F. Bruinsma, W. M. Gelbart:
     Europhys. Lett. {\bf 46} (1999) 454
%
\bibitem{safinya}  I. Koltover, T. Salditt, C. R. Safinya:
     Biophys. J. {\bf 77} (1999) 915
%
\bibitem{levin}  P.S. Kuhn, Y. Levin, M.C. Barbosa:
     Physica A {\bf 274} (1999) 8
%
\bibitem{eugene}  E. Gurovitch, P. Sens:
     Phys. Rev. Lett. {\bf 82} (1999)  339
%
\bibitem{losrusos}  M. H. Panckhurst, Aust. J. Chem. {\bf 15} (1962) 383 \\
     D. R. Rosseinsky: J. Chem. Soc. (1962) 785
%
\bibitem{feynman} An elementary introduction to the Madelung
    constant appears in  R. P. Feynman, R. B. Leighton,
    M. Sands: {\it The Feynman lectures in Physics},
    Section II.8.3, Reading: Addison-Wesley 1964
%
\bibitem{fisher}  M.E. Fisher and Y. Levin:
     Phys. Rev. Lett. {\bf 71} (1993) 3826
%
\bibitem{inprepa}  F. J. Solis, M. Olvera de la Cruz: unpublished
%
%
\bibitem{delsanti}  I. Sabbagh, M. Delsanti, P. Lisieur:
     Eur. Phys. J. B {\bf 12} (1999) 253
%
\bibitem{amis}  B.D. Ermi, E. J. Amis:
    Macromolecules {\bf 31} (1998) 7378
%
\bibitem{claudine}  W. Essafi, F. Lafuma, C.E. Williams:
    Eur. Phys. J. B {\bf 9} (1999) 261
%
%
\end{thebibliography}
\end{document}